\documentclass[preprint]{aastex}
\usepackage{natbib}
\usepackage{graphicx,times,amssymb}
\shorttitle{Planet Hunters : New \textit{Kepler} quarter 2 planet candidates}
\shortauthors{Lintott et al.}

\begin{abstract}
We present new planet candidates identified in NASA \textit{Kepler} quarter two public release data by volunteers engaged in the Planet Hunters citizen science project. The two candidates presented here survive checks for false-positives, including examination of the pixel offset to constrain the possibility of a background eclipsing binary. The orbital periods of the planet candidates are 97.46 days (KIC~4552729) and 284.03 (KIC~10005758) days and the modeled planet radii are 5.3 and 3.8 R$_{\oplus}$. The latter star has an additional known planet candidate with a radius of 5.05 R$_{\oplus}$ and a period of 134.49 days which was detected by the \textit{Kepler} pipeline. The discovery of these candidates illustrates the value of massively distributed volunteer review of the \textit{Kepler} database to recover candidates which were otherwise uncatalogued. 
\end{abstract}

\keywords{planetary systems -- stars: individual (KIC~4552729, KIC~10005758)}

\begin{document}

 \title{Planet Hunters: New \textit{Kepler} planet candidates from analysis of quarter 2\thanks{This publication has been made possible by the participation of more than 100,000 volunteers in the Planet Hunters project. Their contributions are individually acknowledged at \texttt{http://www.planethunters.org/authors}}}

 \author{Chris J. Lintott\altaffilmark{2,3}, Megan E. Schwamb\altaffilmark{4,5,6}, Thomas Barclay\altaffilmark{8,9}, Charlie Sharzer\altaffilmark{7}, Debra A. Fischer\altaffilmark{7}, John Brewer\altaffilmark{7}, Matthew Giguere\altaffilmark{7}, Stuart Lynn\altaffilmark{3}, Michael Parrish\altaffilmark{3}, Natalie Batalha\altaffilmark{8}, Steve Bryson\altaffilmark{8}, Jon Jenkins\altaffilmark{11}, Darin Ragozzine\altaffilmark{10}, Jason F. Rowe\altaffilmark{8}, Kevin Schwainski\altaffilmark{4,5}, Robert Gagliano\altaffilmark{12}, Joe Gilardi\altaffilmark{12}, Kian J. Jek\altaffilmark{12}, Jari-Pekka P\"a\"akk\"onen\altaffilmark{12}, Tjapko Smits\altaffilmark{12}}
 
 \altaffiltext{2}{cjl@astro.ox.ac.uk; Oxford Astrophysics, Denys Wilkinson Building, Keble Road, Oxford OX1 3RH}
 \altaffiltext{3}{Adler Planetarium, 1300 S. Lake Shore Drive, Chicago, IL 60605, USA}
 \altaffiltext{4}{Department of Physics, Yale University, P.O. Box 208121, New Haven, CT 06520, USA}
 \altaffiltext{5}{Yale Center for Astronomy and Astrophysics, Yale University, P.O. Box 208121, New Haven, CT 06520, USA}
 \altaffiltext{6}{NSF Fellow}
 \altaffiltext{7}{Department of Astronomy, Yale University, New Haven, CT 06511 USA}
 \altaffiltext{8}{NASA Ames Research Center, Moffett Field, CA 94035, USA}
\altaffiltext{9}{Bay Area Environmental Research Institute, 560 Third St West, Sonoma, CA 95476, USA}
 \altaffiltext{10}{Harvard-Smithsonian Center for Astrophysics, 60 Garden Street, Cambridge, MA 02138, USA}
 \altaffiltext{11}{SETI Institute, Mountain View}
 \altaffiltext{12}{Planet Hunters}



\maketitle

\label{firstpage}

\section{Introduction}

The \textit{Kepler} mission \citep{Borucki} has made a remarkable contribution to our knowledge of the population statistics of planets beyond our solar system. Building on the legacy of ground-based planet searches that have utilised techniques including Doppler observations, transit photometry, microlensing and direct imaging to identify more than 700 exoplanets \citep{schneider11, wri11}, more than 2000 planet candidates have now been announced by the \textit{Kepler} team \citep{NewKep}. These candidates are the result of monitoring more than 150,000 stars with a rapid, 29.4 minute, observing cadence with excellent photometric precision, approaching 30 ppm \citep{Gilliland}. Data from the mission are being released to a public archive hosted by the Mikulski Archive for Space Telescopes at STScI (MAST\footnote{http://archive.stsci.edu/}) 
and the NASA Exoplanet Archive\footnote{http://exoplanetarchive.ipac.caltech.edu/}. This paper reports the discovery of additional candidates made by visual inspection of the first four months of Kepler data by volunteer citizen scientists using the `Planet Hunters'\footnote{http://www.planethunters.org} interface. 

The Planet Hunters website is one of several citizen science projects to make use of the Zooniverse\footnote{http://www.zooniverse.org} platform, first described in \citet{lin08} and \citet{Smithsn}. Launched on 2010 December 16 and subsequently updated as described below, the site presents volunteers with data extracted from the Kepler archive, asking them to mark features which appear transit-like. Such classifications draw on the intrinsically human ability for pattern recognition, recognising features of interest  despite the possible presence of glitches or other artifacts that would affect less-flexible machine learning approaches to the problem. While inspection of even a fraction of the \textit{Kepler} dataset by a small number of experts would be prohibitively time consuming, by sharing the task between tens of thousands of volunteers Planet Hunters makes large-scale visual classification possible. Having multiple independent classifications available in each case is also a significant advantage of this technique. The Planet Hunters website is available in English and in Polish\footnote{Polish translations were provided by a team led by L. Mankiewicz and J. Pomierny }. 

The Planet Hunters approach is thus complimentary to that taken by the \textit{Kepler} team, who have developed the Transit Planet Search (TPS) algorithm, a wavelet-based adaptive filter to identify a periodic pulse train with temporal widths ranging from 1 to 16 hours \citep{jenkins02, jenkins10}. Photometric uncertainties are assessed to identify light curves with phase-folded detection statistics exceeding a significance threshold and periods greater than 12 hours \citep{Tenenbaum}. Additional data validation including further automatic light-curve fitting is then carried out, followed by visual review of likely candidates before they are identified as planet candidates, or \textit{Kepler} objects of interest (KOI) \citep{bat10}. The thousands of discoveries made using these automatic routines are testament to their effectiveness, but is in important to note that independent review using methods with different biases and sensitivities, such as that provided here, are important. The fact that candidates of interest are visually inspected by the \textit{Kepler} team further underscores the importance of visual inspection in planet discovery. 

The first two Planet Hunters candidates were reported by \citet{PH1} following analysis of classifications made during the first month of the site's operation. Following the framework presented in \citet{mj11}, the false positive probabilities for these candidates were 0.3\% and 5.0\%, low enough to present confident detections of candidates. While these candidates were found via simple inspection of classifications provided by volunteers, \citet{Schwamb} have since undertaken a systematic analysis of candidates in order to measure efficiency using synthetic planets inserted into the data and the Kepler sample of planets with periods less than 15 days. Although performance drops rapidly for smaller radii, reaching 40$\%$ for $2-3{R_{\oplus}}$, above $4{R_{\oplus}}$ Planet Hunters is better than 85$\%$ efficient at identifying transits. 

\section{Identifying Transits}

Kepler data from the first quarter were released in 2010 June, followed by the second quarter in 2011 February, and these two data sets formed the basis of  the search reported in this paper. The `quarter 1' data release consists of the first 33.5 days of \textit{Kepler} science data, while `quarter 2' contains 93 days of observations from 2009 May 1 to 2009 September 17. Data quality varied between the quarters, with quarter 2 data suffering from a greater number of artifacts. These were primarily due to sudden changes in brightness caused by sudden intentional changes in pointing performed to compensate for the larger than expected pointing drift experienced during quarter 2 \citet{KDCH}. The effect of this change in data quality on detection efficiency will be discussed in a later paper. 

The Planet Hunters interface was described in \citet{PH1} and \citet{Schwamb}. Following the viewing of an extremely short in-line tutorial, volunteers were presented with a randomly selected light curve from initially quarter 1 and then both quarters 1 and 2 of \textit{Kepler} data. In order to present data from both quarters in a consistent manner, quarter 2 data is split into sections of 30 days, each with a 5 day overlap, so that nothing is lost between segments. After answering a small number of questions, used in more systematic searches of the data, the volunteer can then mark the position of one or more transits. After at least five classifications the light curve was withdrawn from classification; in total, between 5 and 10 classifications are obtained for each light curve. 

\section{Kepler Planet Hunters Candidates} 

\begin{figure}
\includegraphics[width=0.98\textwidth,angle=0]{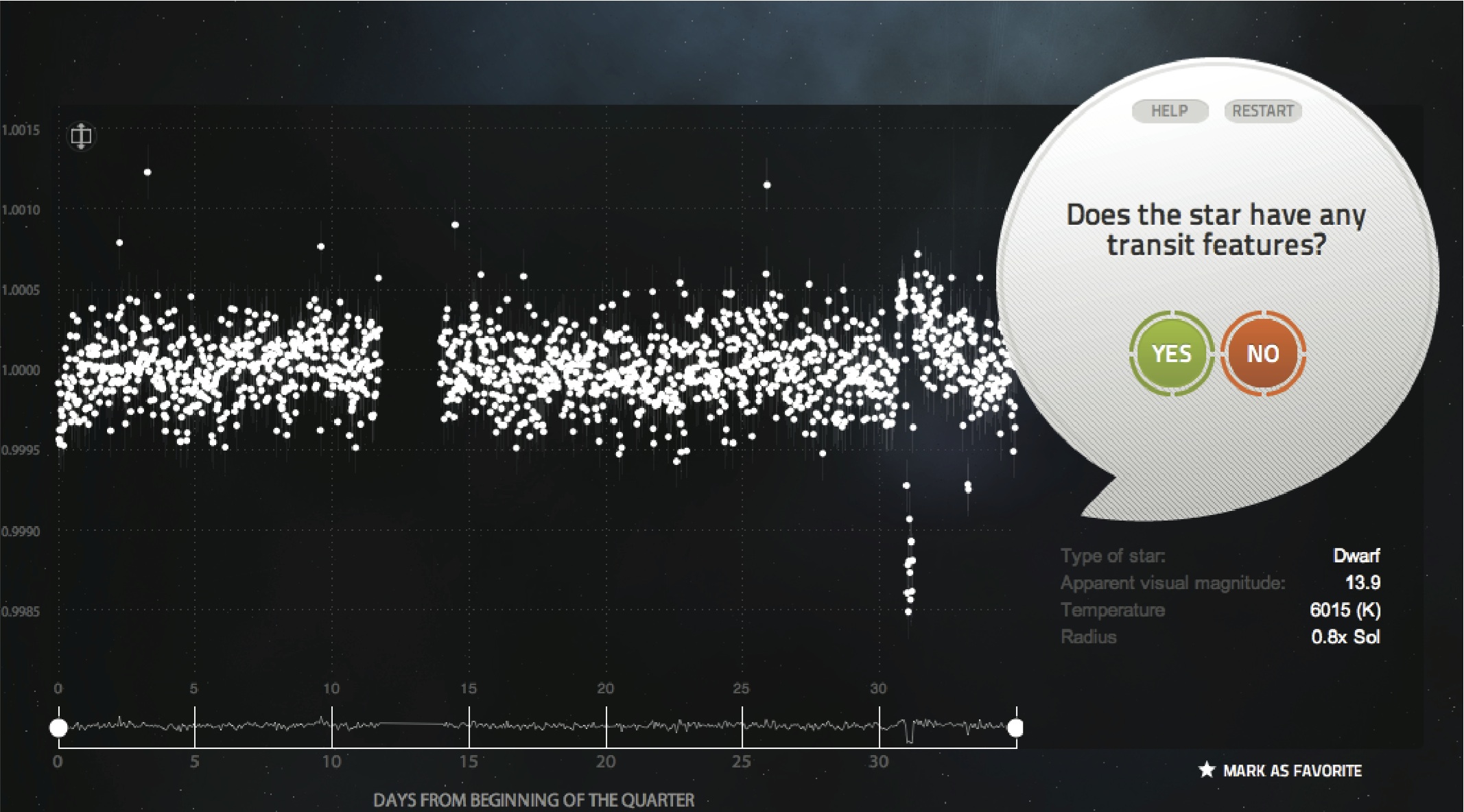}
\caption{Quarter 2 data for KIC10005758 viewed in the Planet Hunters interface}\label{fig:interface}
\end{figure}

The latest release of \textit{Kepler} candidates, from the first sixteen months of data, includes nine systems in which planet candidates were independently discovered by Planet Hunters \citep{NewKep}. In this paper, we report on two additional new candidates which were discovered by Planet Hunters but not by the procedures used by the  \textit{Kepler} team. These candidates were initially drawn to the attention of the science team by posts on the `Talk' section of the Planet Hunters website\footnote{http://talk.planethunters.org. The code is available under an open-source license at https://github.com/zooniverse/Talk}. Talk is an object-orientated discussion tool which is integrated with the classification process, designed to enable discussion about objects of interest which can then be easily brought to the attention of the science team. In particular, Lubomir Stiak collected likely candidates from posts by other users, particularly Kian Jek and Jari Paakkonen for the two candidates discussed here. Each candidate was also viewed in the main Planet Hunters interface (see figure \ref{fig:interface}) by multiple volunteers who successfully marked the relevant transits. Along with other likely transits, these were passed to our Kepler co-authors who were able to examine the light curves with their data verification pipeline, confirming that the objects discussed here were likely planet candidates. 

In addition to the two sources discussed below, transit-like events were identified in quarters 1 and 2 for the stars KIC 3326377, 5511081, 6504954, 7761918, 8160953, 11875734, 6268648, 5864975. Additional transit-like events were identified in KIC 5371776. These candidates had already independently been identified by the \textit{Kepler} team and are included in the \citet{NewKep} catalogue. 

\subsection{KIC 4552729}
\begin{figure}
\includegraphics[width=0.88\textwidth,angle=0]{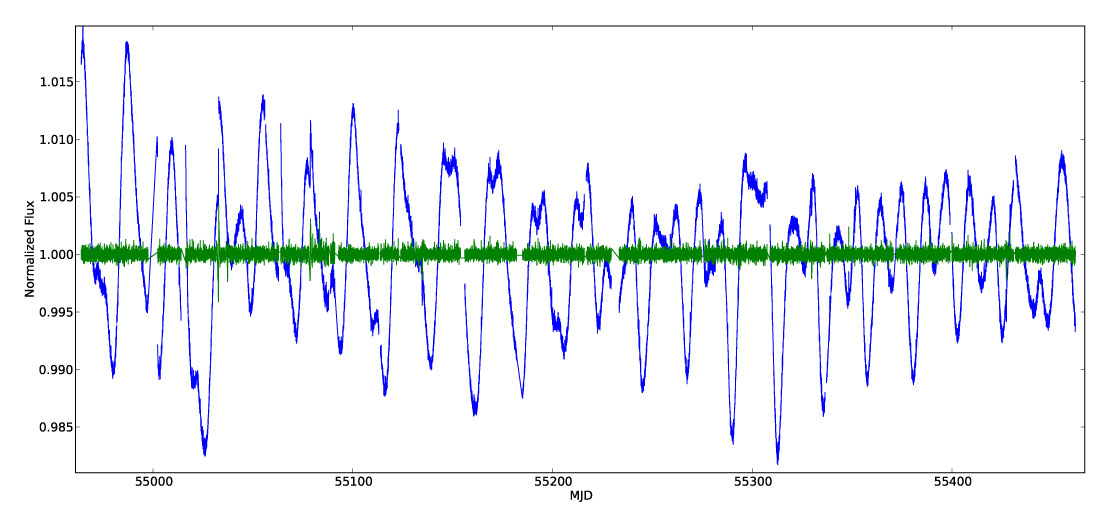}
\caption{The light curve from quarter 1 to 6 for KIC4552729, after correction for instrumental effects, is shown in blue. The smoothed curve, shown in more detail in Figure \ref{fig:4552729}, is shown here in green.}\label{fig:4552729_unsmooth}
\end{figure}

\begin{figure}
\includegraphics[width=0.88\textwidth,angle=0]{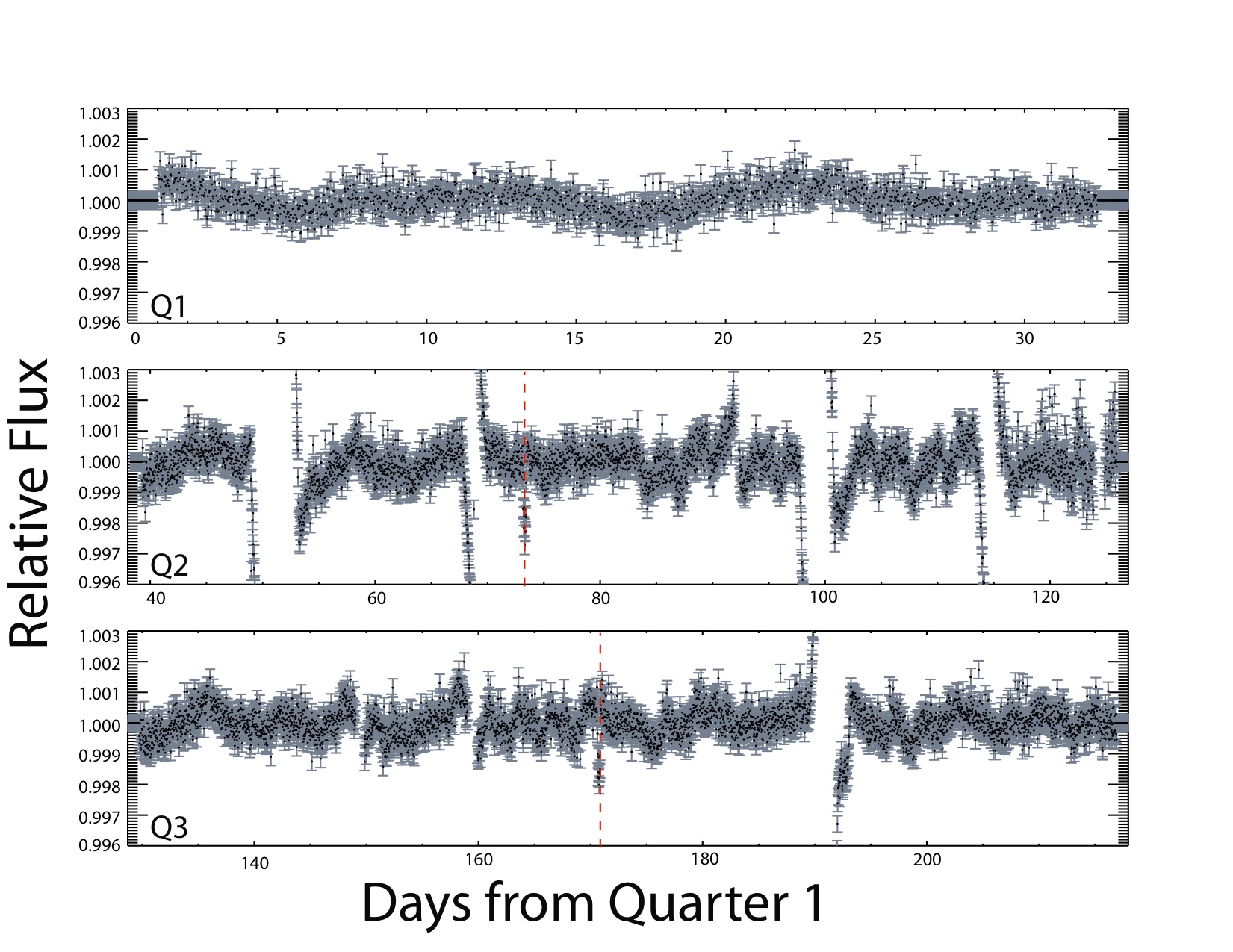}
\includegraphics[width=0.88\textwidth,angle=0]{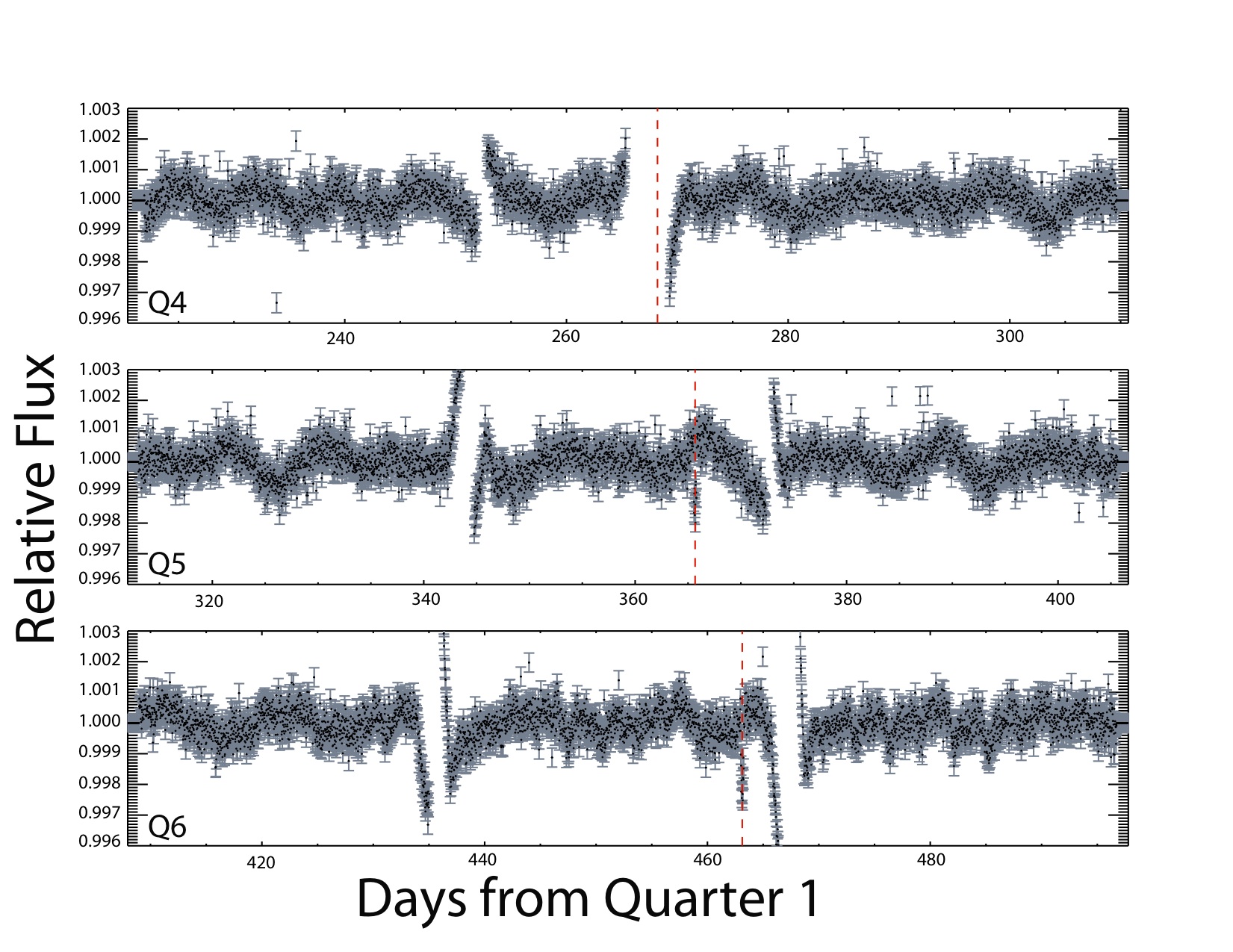}
\caption{Light curves from quarters 1 to 6 for KIC 4552729, with transits identified by a solid red line. Large amplitude variability has been removed by the dividing out of a smoothed curve, produced via the application of a box car filter with a width of 2.33 days.} \label{fig:4552729}
\end{figure}

KIC 4552729 is listed in the Kepler Input Catalog \citep{kmt09} with a magnitude of 14.98 and a g-r colour of 0.982. $\textrm{T}_{\textrm{eff}}$ is 4620 K, $\log \left(\textrm{g}\right)$ is 4.390, [Fe/H] is 0.267 solar and R=$0.977\textrm{R}_{\odot}$.  The light curve for this star for quarters 1-6 of Kepler data is shown in Figure \ref{fig:4552729_unsmooth}, showing significant variability. A Lomb-Scargle periodogram \citep{Scargle} was performed and a stellar rotation rate of 22.4 days for KIC 4552726 was derived. At this point, a boxcar filter with a width of 2.33 days was used to remove large amplitude variability, and the resulting light curve is shown in Figure \ref{fig:4552729}.

A single transit was first identified in quarter 2 via the Planet Hunters interface, initially by Robert Gagliano and then by seven other Planet Hunters volunteers.  Inspection of additional data reveals that transit events with a depth of approximately 0.3\% are clearly seen in quarters 2,3,5 and 6, repeating with a 97.5 day period. An expected transit in quarter 4 fell in a gap in observations. While transits in quarters 3, 5 and 6 were identified by TPS, the star did not pass the \textit{Kepler} data verification process due to the presence of large systematics in the data (Batalha, private communication) which affected the fit, preventing the source from being passed for further feedback. More specifically, the presence of these systematics resulted in statistics being returned for an erroneous period. Upgrades to the pipeline carried out since the discovery have improved outlier detection, and KIC 4552729 is now detected by the pipeline (Jenkins, private communication). In addition, the shape and depth of the events seem consistent with the presence of a planet. 

\begin{figure}
\includegraphics[width=0.88\textwidth]{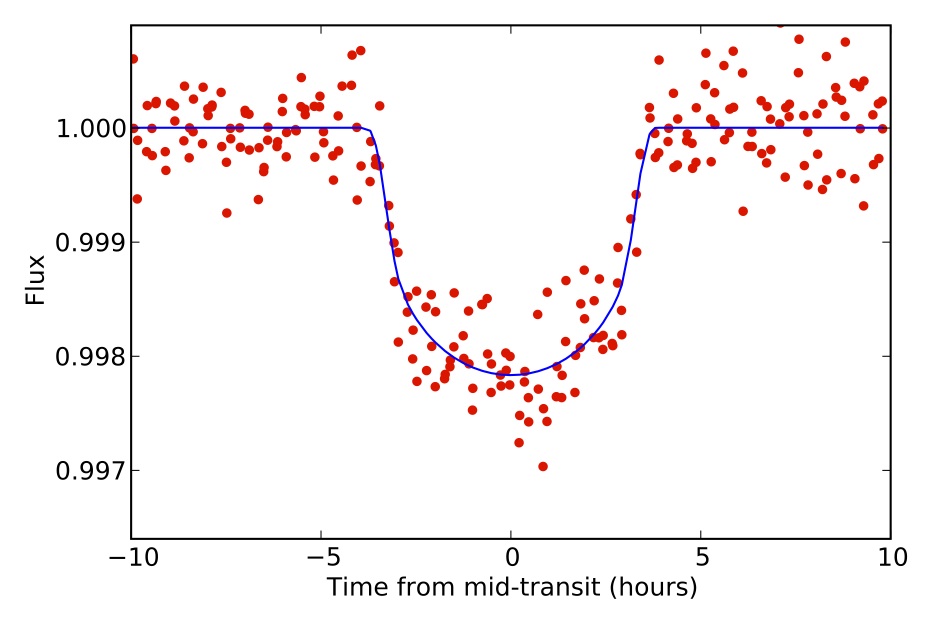}
\caption{Fit to the folded light curve for KIC4552729}\label{fig:fits1stcand}
\end{figure}

A least squares fit was made to the data, using the \citet{MandelAlgol} transit model and non-linear limb darkening parameters from \citet{Claret}. The mean stellar density was a free parameter and a circular model was assumed. Assuming the stellar parameters given above are accurate (despite potentially errors as large as $\sim 50$\% \citep{Verner, kmt09}), and based on all observed transits, we obtain an orbital period of $97.45502\pm 0.00094$ days and an impact parameter of $0.765 \pm 0.037$. The derived planetary radius is $0.0502 \pm 0.0028 \mathrm{R_{*}}$, or $5.3 \mathrm{R_{\oplus}}$. The fit is illustrated in figure \ref{fig:fits1stcand}. The mean stellar density of this fit $0.238\pm 0.072~ \mathrm{g cm^{-3}}$, which is low compared to what might be expected for a star of the given temperature and surface gravity which might indicate a (sub)giant, although the KIC radius is too small for such a star. Further spectroscopy would help in constraining the mean stellar density and hence the fit. Similar results were obtained when a curve was fitted to the data using a Levenberg-Marquardt least-squares fitting routine, using the MPFIT IDL routine \citep{MPFIT}. 

\begin{figure}
 \includegraphics[width=0.88\textwidth]{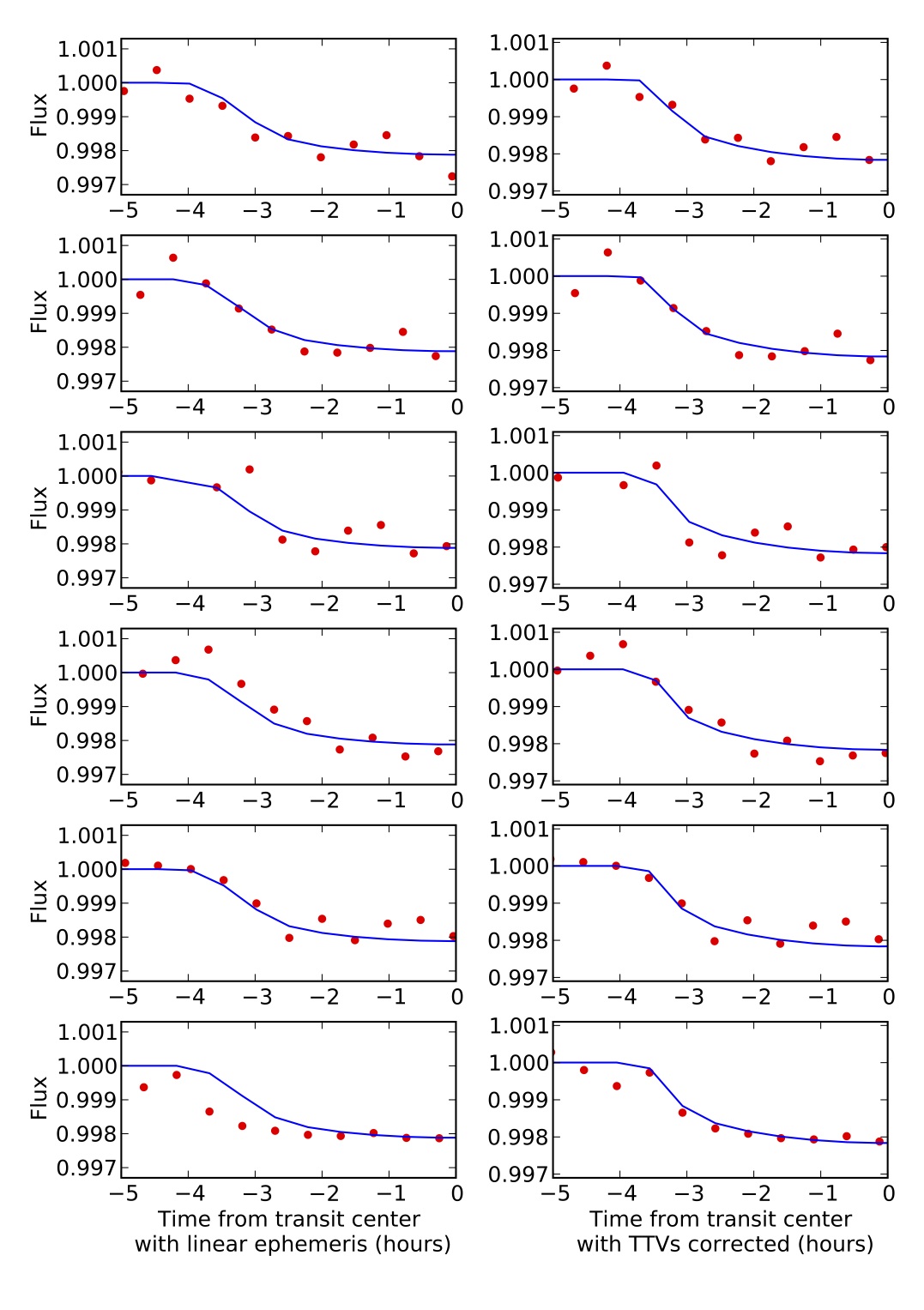}
 \caption{Left : The ingress for each of the six transits observed for KIC4552729. A clear offset from a linear ephemeris at times of around -1 hours is seen for many of the transits. Right : The effect of the correction applied to transit timing in order to improve the planet properties. In all panels the blue line is the best fit transit model after a correction for the presence of transit timing variations has been applied.}\label{fig:TTV}
\end{figure}

However, there are indications of significant transit timing variations (TTVs) with an approximate amplitude of an hour, which indicate the presence of another mass in the system.  These TTVs are illustrated in figure \ref{fig:TTV}, which shows the ingress for each of the six transits observed for this source. As a full cycle of TTVs is not observed in the currently available public data, the mass of this companion is unconstrained. The effect of these TTVs into increase the apparent scatter and hence the errors on transit measurements, but they may be removed by slightly adjusting the time of each observation in order to force a linear transit ephemeris, as shown in the figure. The last observed transit shows the greatest offset from predicted transit times. However, removing the TTVs to produce a linear ephemeris the fit improves, and we obtain an orbital period of $97.45530\pm0.00094$ and an impact parameter of $0.329\pm0.037$. Critically, the mean stellar density obtained is $1.545\pm0.072  \mathrm{g cm^{-3}}$, more consistent with the known stellar properties. The derived planetary radius is $0.0436\pm 0.0013 \mathrm{R_{*}}$, or $4.65 \mathrm{R_{\oplus}}$.
 
Independent measurements of the mass of the transiting object are not available, and so KIC 4552729 is a planet candidate rather than a confirmed planet. The possibility remains that the transits are due not to a planet, but to the presence of a background eclipsing binary. The blending of light from such a system with that from the brighter foreground star can result in a composite light curve with a sufficiently shallow transit depth to cause confusion. 

A common method of constraining the presence of a background eclipsing binary is to examine any offset in position between stacked images taken in and out of transit. If the transit is occurring on the central source, rather than on a background contaminant, no net apparent motion is expected. If a background binary is responsible, then the changing brightness ratio between foreground and background star will result in an apparent centroid shift. The average multi-quarter offset from quarters 1 to 8 for this candidate is 1.5$\sigma$, providing support for a planetary origin for the transits. 

\subsection{KIC 10005758}

\begin{figure}
\includegraphics[width=0.88\textwidth,angle=0]{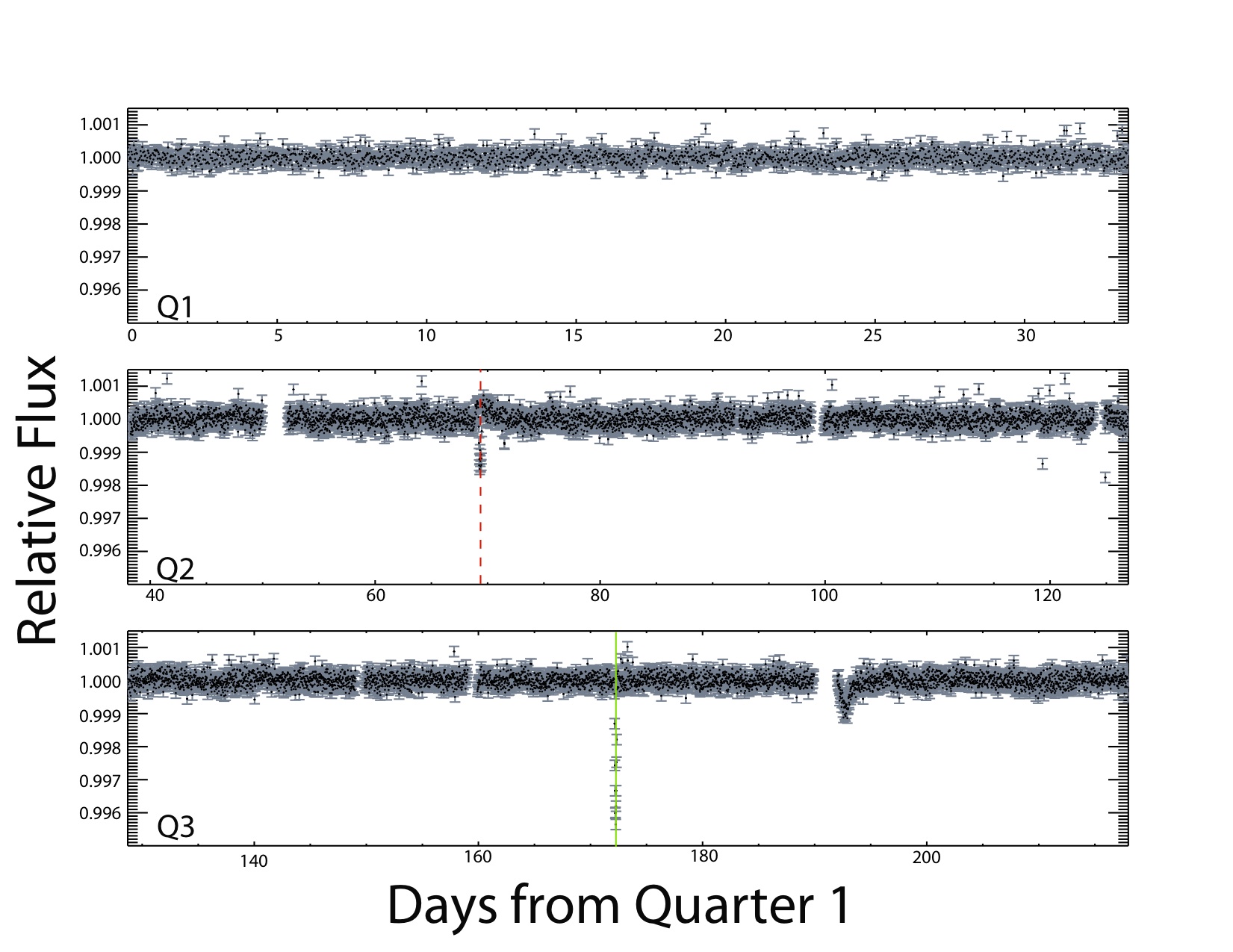}
\includegraphics[width=0.88\textwidth,angle=0]{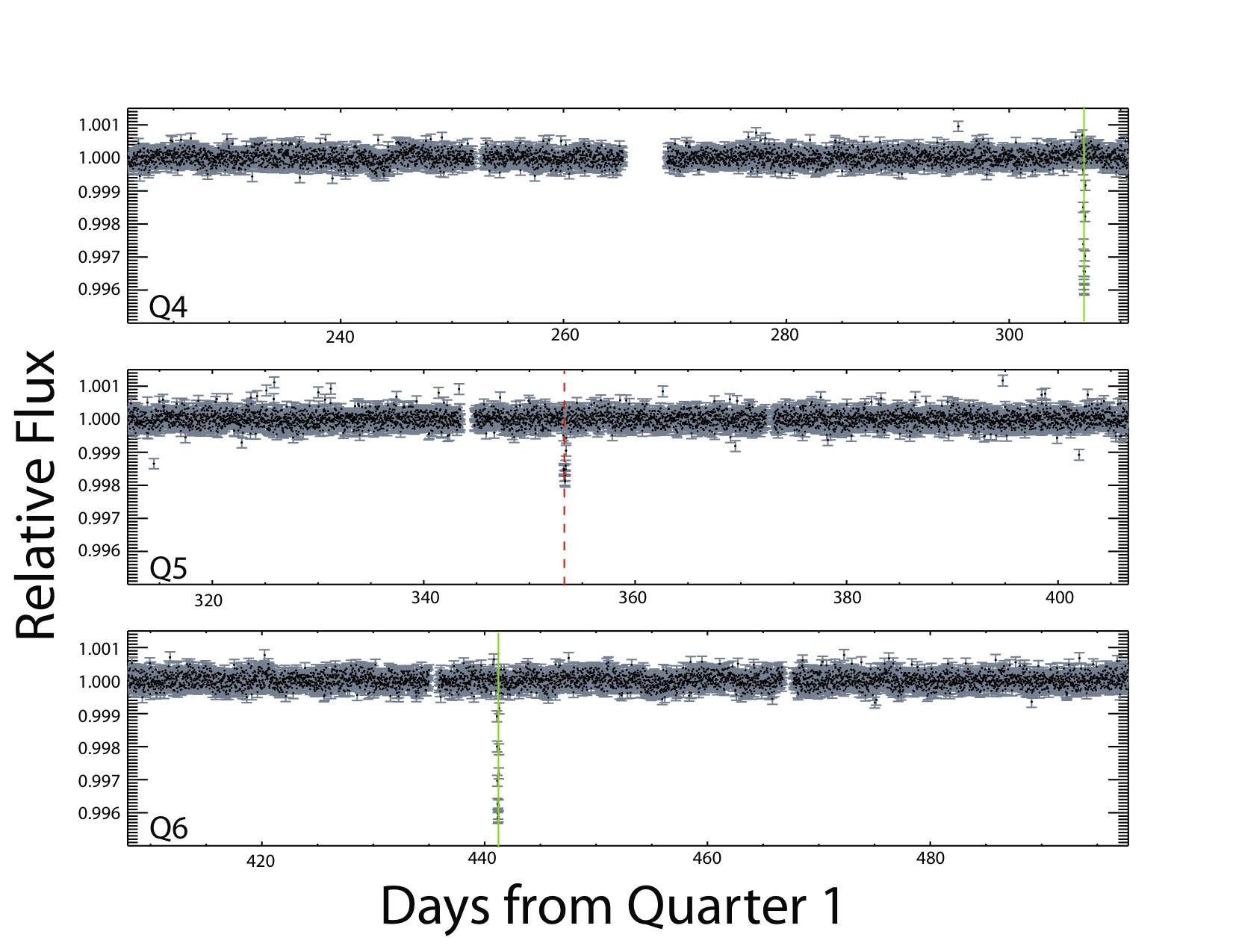}
\caption{Light curves for quarters 1 to 6 for KIC 10005758, with transits belonging to the series identified by Planet Hunters marked by a dashed red line and those from the series identified by the \textit{Kepler} team by a solid green line.} \label{fig:10005758}
\end{figure}

KIC 10005758 (KOI 1783) is listed in the Kepler Input Catalog \citep{kmt09} with a magnitude of 13.9 and a g-r colour of 0.410. $\textrm{T}_{\textrm{eff}}$ is 6015 K, $\log (\textrm{g})$ is 4.692, [Fe/H] is -0.247 solar and R=$0.766\textrm{R}_{\odot}$. The catalogue presented by \citet{NewKep} includes an increased value for the radius, R=$0.930\textrm{R}_{\odot}$ and a new value from $\log\left(\textrm{g}\right)=4.151$ and these new values is used throughout our analysis. The light curve for the first six quarters of \textit{Kepler} data is shown in figure \ref{fig:10005758} and shows two separate sets of transits. Planet Hunters volunteers, initially Joe Gilardi and Tjapko Smits, identified the transit visible in quarter 2\footnote{The transit lay in a region which was shown in two separate Planet Hunters lightcurves}, which took place roughly 69.3 days after the beginning of quarter 1. Inspection of data from quarters 4-6 data, made public in January 2012, show that this transit signature, with an depth of 1666 ppm, is repeated 284 days later. 

While each individual transit was detected by TPS, this set of transits was not identified by the \textit{Kepler} review process used by the team (Batalha, Jenkins, private communication) and the system was not promoted to KOI status. The \textit{Kepler} team did identify a second series of transits with a shorter period, with a depth of 3801 ppm which is first seen 172 days after the beginning of Quarter 1, and repeats with a roughly 132 day period. The first of this series falls between the period covered by quarter 1 and that covered by quarter 2, and was thus not visible in the data. The system is included in the list of `V' shaped candidates given as table 2 in \citet{NewKep}. Such a transit profile might indicate that the events are due to an eclipsing binary, or a grazing planet transit; shape alone is thus not sufficient to distinguish stellar from planetary transits. KIC 10005758 is identified in \citet{NewKep} as a single candidate system, which our fit indicates has a radius of R=6.606R$_{\oplus}$.

\begin{figure}
\includegraphics[width=0.88\textwidth]{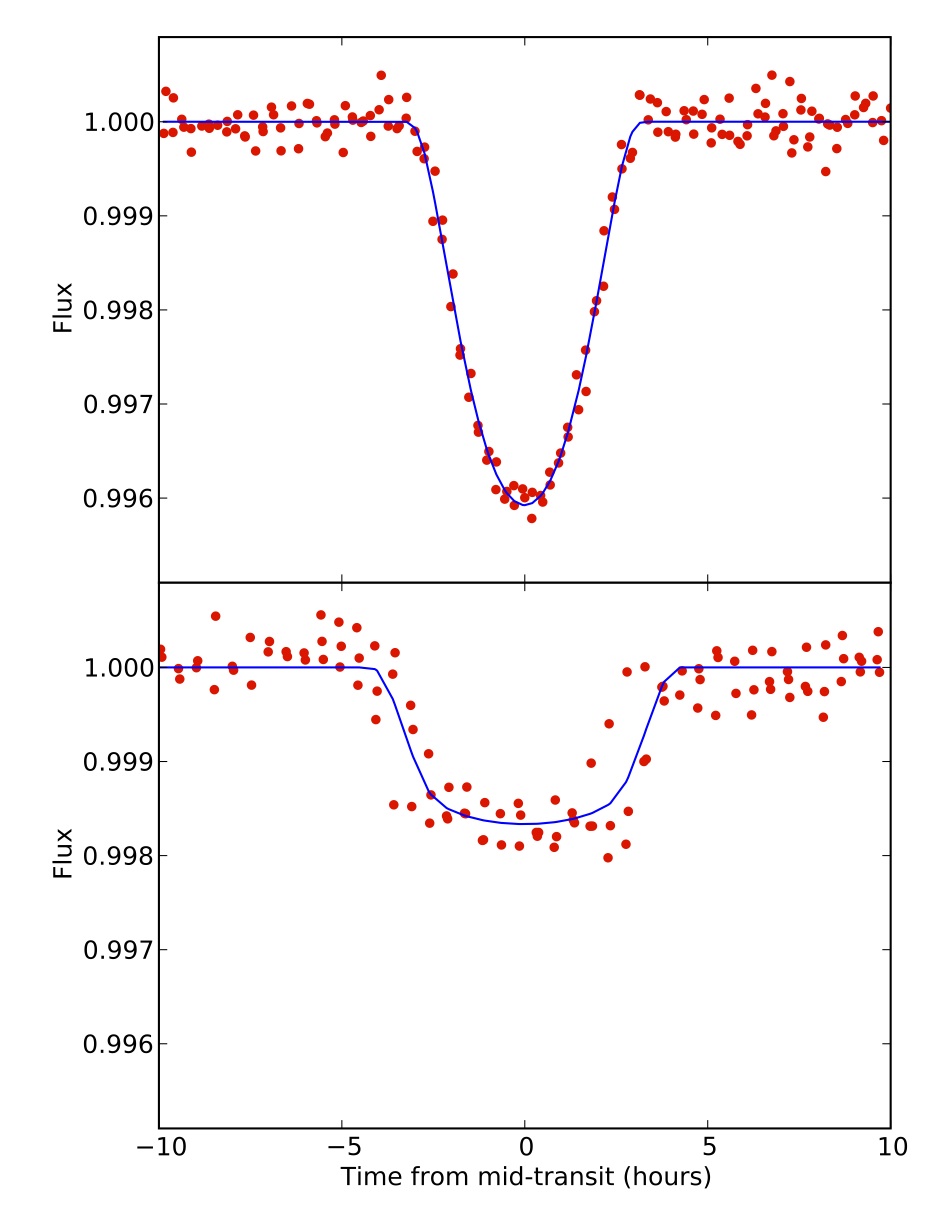}
\caption{Fit to the folded light curve for KIC10005758b (top) and c (bottom), the latter the candidate identified by Planet Hunters volunteers.}\label{fig:fits2ndcand}
\end{figure}

 Fitting was carried out as before, and is illustrated in Figure \ref{fig:fits2ndcand}. The mean stellar density derived from the fit is $0.742\pm 0.071\mathrm{g cm^{-3}}$ and other derived parameters are given in table \ref{tab:2ndcan}. The two planet candidates are fitted simultaneously, assuming circular orbits. 
 
The detection of two separate series of transits dramatically lowers the chance of contamination by background eclipsing binaries, requiring either two binaries or contamination from an eclipsing binary and a star with a transiting planet in order to account for both sets of transits. We follow the procedure used by the Kepler team in subtracting stacked in-transit images from stacked out-of-transit images, and checking for an offset in centroid position between this subtracted image and the stacked out-of-transit images. A large offset indicates that the transit is on the background candidate. The average multi-quarter offset in pixel position reported for this candidate is only 0.8$\sigma$, compared to a threshold of 3$\sigma$ for KOIs reported by the \textit{Kepler} team \citep{NewKep}. This supports the interpretation, initially made visually, that this is not a background eclipsing binary but rather an additional planet candidate. KIC 10005758 is therefore likely to be a multi-planet system.

\begin{table}
\begin{tabular}{| l | c | c |}
\hline
Candidate & KIC~10005758b & KIC~10005758c\\
\hline
Orbital Period (days) & $134.47975\pm 0.00060$ & $284.0433\pm0.0022$\\
Impact Parameter & $0.951 \pm 0.019$ & $0.900\pm0.016$\\
Planet Radius ($\mathrm{R_{*}}$) & $0.0783\pm0.0038$ & $0.04412\pm0.00071$\\
Planet Radius ($\mathrm{R_{\oplus}}$)& 8.012 & 4.596\\
\hline 
\end{tabular}
\caption{Derived parameters for candidates associated with KIC10005758, using stellar parameters in \citet{NewKep}. KIC10005758b was identified by the \textit{Kepler} team, and KIC10005758c by Planet Hunters volunteers} \label{tab:2ndcan}
\end{table}

The presence of two planet candidates make this one of hundreds of Kepler systems with multiple transiting planets \citep{NewKep}. As a doubly-transiting system, this candidate is more valuable than a star with only one transiting planet \citep{Ragozzine}, at least in part because multi-transiting systems are more robust and reliable \citep{Latham}. Estimates of the false positive frequency in doubly-transiting systems suggest that the probability of two false positives is very low, because each false positive is a rare and independently random event \citep{Lissauer}. The apparent TTVs noted above for KIC 4552729 suggest that this argument can be used to support that claim, but the presence of two candidates in the KIC 10005758 system increase the likelihood that they are true planets by a factor of $\sim$10, with the main false positive mode being one true planet around the target star and one background eclipsing binary or a planet around another star \citep{Lissauer12}.  In some cases, a near-resonant ratio of periods is an additional strong indicator that two candidates are planets in the same system, but the period ratio of 2.11 does not provide a compelling likelihood boost. 

\citet{Lissauer} note that a pair of planets initially on coplanar circular orbits will be stable if they can never develop crossing orbits, which will be true when the following condition is met :

\begin{equation}
\Delta = \frac{a_0-a_i}{R_H} > 2\sqrt{3}
\end{equation} 

where $\mathrm{R_H}$ is the radius of the mutual Hill sphere for the two planets. For the system discussed here, $\Delta=9.4$ and so the system meets this condition for orbital stability. 

\section{Discussion} 

We have presented evidence which supports the discovery of two new planet candidates around the stars KIC 4552729 and KIC 10005758 by volunteers using the Planet Hunters citizen science interface. As with many \textit{Kepler} candidates, they are unsuitable for ground-based confirmation via radial velocity measurements, with a signal much smaller than the 4-5 $\mathrm{ms^{-1}}$ errors typically obtained. However, the presence of these candidates indicate that visual inspection can provide a valuable additional and complementary technique to the combination of algorithm and expert review used by the Kepler team. As with the two Planet Hunters discoveries reported by \citet{PH1}, the two candidates presented here were identified by the Transit Planet Search algorithm as threshold crossing events but were not promoted to candidate of KOI status. 

By assuming an albedo of the planet, it is possible to estimate the equilibrium temperature for the two planet candidates.  Following the prescription of \citet{Borucki11} we calculate the equilibrium temperature for a spherical airless gray body assuming a Bond albedo of 0.3 and emissivity of 0.9. We assume the body is rotating sufficiently fast that the body evenly reradiates the absorbed stellar flux.  Using the planet orbital period and stellar density derived from the fits described above, and assuming the KIC reported stellar temperature, we find an equilibrium temperature of 320K and 330K for the planets discovered by volunteers around  KIC 4552729 and KIC 10005758 respectively. If, alternatively, the values for the host star's $\log \left(\textrm{g}\right)$, stellar radius, and temperature are taken from the Kepler Input catalog for KIC 4552729 and from the updated \citet{NewKep} for KIC 10005758, we obtain an equilibrium temperature of 429K and 285K respectively for the  KIC 4552729 and KIC 10005758 candidates. If we adopt the greenhouse warming typical of an Earth-like atmosphere (33K), surface temperatures would be 353K for KIC 4552729 and 363K for KIC 10005758 (462K and 318K respectively assuming catalogue stellar parameters), placing both our discoveries potentially in the habitable zone of their parent stars. Although in each case the planets are likely to be too large to have rocky surfaces, the possibility of habitable moons with rocky or watery surfaces remains. 

The performance of the TPS system continues to be excellent, as shown by the large number of candidates in \citet{NewKep}, many of which were independently recovered by analysis of the Planet Hunters results. However, the presence of identifiable and viable candidates recovered only by pure visual inspection continues to support the use of this novel and complementary technique, particularly as analysis of planet populations which much make assumptions about the completeness of \textit{Kepler} candidates becomes more common. Although the candidates presented here and in \citet{PH1} were not missed through human error, as visual inspection also plays a role in the \textit{Kepler} analysis process, it is perhaps significant to note that review by a large number of inexpert reviewers can, with appropriate weighting, be more accurate than review by a small number of experts \citep{lin08}. Planet Hunters is essentially a sensitive search for single transits, which can then be used to flag datasets of especial interest for further examination. This ability is particularly important for systems where features such as significant transit timing variations make routines based on fitting multiple transits ineffective \citep{2011MNRAS.417L..16G}.

\acknowledgements{The data presented in this paper are the result of the efforts of the Planet Hunters volunteers, without whom this work 
would not have been possible. In addition to those volunteers listed as authors, the following list of people flagged transit events for the light curves discussed in this 
paper: Pamela Fitch, Dr Johann Sejpka, Gregoire P.A. Boscher, Matthew Lysne, Thanos Koukoulis and Andre Engels (KIC 4552729); Ben Myers, Daniel Posner, Terrence Goodwin, Theron Warlick, Charles Bell, David Lindberg, Sean Parkinson, Samuel Randall, Eduardo Mari\~no, Frank Barnet, Terrence Goodwin, Ewa Tyc-Karpinska,  Heinz W. Edelmann, Lynn van Rooijen-McCullough, Gary Duffy, `kamil', Branislav Marz, `Adnyre',  Colin Pennycuick (KIC 10005758)

DF acknowledges funding support from Yale University and support from the 
NASA Supplemental Outreach Award, 10-OUTRCH.210-0001. MES is supported by an NSF Astronomy and Astrophysics Postdoctoral Fellowship under award AST-100325. The Zooniverse is supported by The Leverhulme Trust. The Talk system used by Planet Hunters was built during work supported by the National Science Foundation under Grant No. DRL-0941610.
We gratefully acknowledge the dedication and achievements of Kepler Science Team 
and all those who contributed to the success of the mission. We acknowledge use of public release data 
served by the NASA/IPAC/NExScI Star and Exoplanet Database, which is operated by the Jet Propulsion 
Laboratory, California Institute of Technology, under contract with the National Aeronautics and Space Administration.  
This research has made use of NASA's Astrophysics Data System 
Bibliographic Services. This paper includes data collected by the \emph{Kepler} spacecraft, and we gratefully acknowledge the entire \emph{Kepler} mission team's efforts in obtaining and providing the light curves used in this analysis. Funding for the \emph{Kepler} mission is provided by the NASA Science Mission directorate. The publicly released \emph{Kepler} light curves were obtained from the Mikulski Archive for Space Telescopes at the Space Telescope Science Institute (MAST). STScI is operated by the Association of Universities for Research in Astronomy, Inc., under NASA contract NAS5-26555. Support for MAST for non-HST data is provided by the NASA Office of Space Science via grant NNX09AF08G and by other grants and contracts.}

\bibliographystyle{mn}

\begin{thebibliography}{74}

\bibitem[Batalha et al.(2010)]{bat10}
Batalha, N. M., et al. 2010, ApJL, 713, 103

\bibitem[Batalha et al.(2012)]{NewKep}
Batalha N., et al., 2012, In Prep

\bibitem[Borucki et al.(2010)]{Borucki}
Borucki W., et al., 2010, Science, 327, 977


\bibitem[Borucki et al.(2011)]{Borucki11}
Borucki W., et al., 2011, ApJ, 736, 19

\bibitem[Brown et al.(2011)]{kmt09}
Brown et al. 2011, \aj, 142, 112

\bibitem[Christiansen et al.(2011)]{KDCH}
Christiansen, J. L, et al. 2011, Kepler Data Characteristics Handbook (KSCI-19040-002)

\bibitem[Claret \& Bloemen (2011)]{Claret}
Claret \& Bloemen 2011, A\&A, 529, 75

\bibitem[Fischer et al.(2012)]{PH1}
Fischer D.A. et al., 2012, \mnras, 419, 2900

\bibitem[{{Garc{\'{\i}}a-Melendo} \&
 {L{\'o}pez-Morales}(2011)}]{2011MNRAS.417L..16G}
{Garc{\'{\i}}a-Melendo}, E. \& {L{\'o}pez-Morales}, M. 2011, \mnras, 417, L16

\bibitem[Gilliland et al.(2011)]{Gilliland}
Gilliland et al. 2011, ApJS, 197, 6

\bibitem[Jenkins(2002)]{jenkins02} 
Jenkins, J.~M.\ 2002, \apj,575, 493

\bibitem[Jenkins et al.(2008)]{jen08} 
Jenkins, J.~M., et al. 2008 \apj, 724, 1108

\bibitem[Jenkins et al.(2010)]{jenkins10} 
Jenkins, J.M. et al.\ 2010, SPIE 7740, 10

\bibitem[Latham et al.(2011)]{Latham}
Latham et al. 2011, \apj, 732, 24 

\bibitem[Lintott et al.(2008)]{lin08}
Lintott, C., et al. 2008, \mnras, 389, 1179

\bibitem[Lissauer et al.(2011)]{Lissauer}
Lissauer et al. 2011, ApJS, 197, 8

\bibitem[Lissauer et al.(2012)]{Lissauer12}
Lissauer et al. 2012, ApJ, 750, 112

\bibitem[Mandel \& Algol(2002)]{MandelAlgol}
Mandel \& Algol 2002, ApJ, 580, 171

\bibitem[Markwardt et al.(2009)]{MPFIT}
Markwardt et al. 2009, ASPC, 411, 251

\bibitem[Morton \& Johnson(2011)]{mj11}
Morton, T. D., Johnson, J. A. 2011, \apj, 738, 170

\bibitem[Ragozzine \& Holman(2010)]{Ragozzine}
Ragozzine \& Holman, 2010, arXiv : 1006.3727

\bibitem[Scargle(1982)]{Scargle}
Scargle, J.D., 1982, ApJ, 263, 835

\bibitem[Schneider(2011)]{schneider11}
Schneider, J. 2011,  http://www.encyclopedia.eu

\bibitem[Schwamb et al.(2012)]{Schwamb} 
Schwamb, M. et al. 2012, accepted by ApJ

\bibitem[Smith et al.(2010)]{Smithsn}
Smith, A. et al., 2011, \mnras, 412, 1309

\bibitem[Tenenbaum et al.(2012)]{Tenenbaum}
Tenenbaum et al. 2012, ApJS, in press, arXiv : 1201.1048

\bibitem[Verner et al.(2011)]{Verner}
Verner, G. A., et al., 2011, \apj, 738, 28

\bibitem[Wright et al.(2011)]{wri11}
Wright, J. T., et al.  PASP, 123, 412

\end{thebibliography}

\clearpage

\label{lastpage}

\end{document}